# Fabrication of superconducting nanowires based on ultra-thin Nb films by means of nanoimprint lithography

Lu Zhao, Yirong Jin, Jie Li, Hui Deng, Hekang Li, Keqiang Huang, Limin Cui and Dongning Zheng

Beijing National Laboratory for Condensed Matter Physics, Institute of Physics, Chinese Academy of Sciences, Beijing 100190, People's Republic of China

E-mail: dzheng@iphy.ac.cn, luzhao2832@163.com

**Abstract:** Nanoimprint lithography (NIL) is an attractive nonconventional lithographic technique in the fabrication of superconducting nanowires for superconducting nanowire single-photon detectors (SNSPDs) with large effective detection areas or multi-element devices consisting of hundreds of SNSPDs, due to its low cost and high throughput. In this work, NIL was used to pattern superconducting nanowires with meander-type structures based on ultra-thin (~4 nm) Nb films deposited by DC-magnetron sputtering at room temperature. A combination of thermal-NIL and UV-NIL was exploited to transfer the meander pattern from the imprint hard mold to Nb films. The hard mold based on Si wafer was defined by e-beam lithography (EBL), which was almost nonexpendable due to the application of IPS® as a soft mold to transfer the pattern to the imprint resist in the NIL process. The specimens fabricated by NIL keep good superconducting properties which are comparable to that by conventional EBL process.

## 1. Introduction

Superconducting nanowire single-photon detector (SNSPD) is a promising candidate for near-infrared photon detection in quantum optics and quantum communications. This type of detector has already demonstrated attractive properties such as picoseconds time resolution, picoseconds jitter time, nearly negligible dark count rate and broad wavelength response (the visible to near-infrared band), which outperform commercially available single-photon detector technologies such as Si and InGaAs avalanche photodiodes (APDs) [1]. The operation principle of the SNSPD is based on a model of a supercurrent-enhanced resistive transition due to the generation of an unstable hotspot in a portion of the quasi-one dimensional superconducting strip [2-4] such as Nb [5, 7], NbN [2, 8], NbTiN [9] and WSi [10]. In order to reach high detection efficiency, the thickness of superconducting layer is generally limited to 10 nm or less. Also, the detector has a meander [1-10] or circular type structure [11] with a narrow strip (100 nm or less) to maximize its active area. Multi-element devices consisting of hundreds of SNSPDs and providing higher counting rates and photon-number resolution capability have been proposed [12]. Thus, ultra-thin superconducting films of high quality and the fabrication of nanowires in larger areas with good uniformity are of great importance for single-photon detectors. Several nanofabrication techniques have been used in SNSPDs fabrication. These include e-beam lithography (EBL) [1-11], focused ion beam (FIB) [13], and local oxidation with an atomic force microscope (AFM) [14, 15]. However, these techniques usually require expensive equipments, and are time consuming for fabricating devices with large active area and, particularly for making the multi-element devices. In the process using FIB, superconducting properties of ultrathin films are often suppressed by the implantation of gallium ions ($Ga^+$) used [16]. Therefore, a low cost and more efficient technique is highly desirable for SNSPDs fabrication.

Among various nanofabrication techniques, nanoimprint lithography (NIL) has attracted more and more interest due to its high-throughput patterning of nanostructures at low cost and high resolution (down to 5 nm) [17-21]. Unlike EBL and FIB, which achieve pattern definition through point by point exposure, NIL is a parallel process lithography technique in which many nanostructures on a film could be patterned simultaneously. An imprint resist is shaped by a NIL mold and consolidation of the resist by controlling temperature (themal-NIL) or UV light exposure (UV-NIL) to replicate patterns. Based on the mechanical deformation of a soft polymer resist, NIL can achieve high resolution without the limitations set by light diffraction or beam scattering that are encountered in conventional lithographic techniques. Whereas the resolution of pattern by NIL relies on the quality of the imprint hard mold defined by conventional lithography such as EBL, FIB or photolithography, which can be used for many times to replicate nanostructures.

NIL technique has stimulated great interest in electronics. Semiconducting devices and organic devices have been made using NIL soon after its invention. However, the application of NIL to the fabrication of superconducting devices has not been reported. In this study we demonstrate the fabrication of superconducting meander nanowires for SNSPDs by means of NIL. A combination of thermal-NIL and UV-NIL was used to transfer the meander-type structures from a NIL hard mold based on Si to ultra-thin (~4 nm) Nb films. During imprint process, the intermediate polymer stamp (IPS®) provided by Obducat was exploited as a soft mold to pattern nanostructures. The meander-type geometries generated based on Nb films were characterized by scanning electron microscopy (SEM). Resistance-temperature (*R-T*) and current-voltage (*I-V*) characteristics of the Nb nanowires were measured. The results showed that nanowires defined by NIL keep good superconducting properties.

2. **Experimental**

*2.1. Deposition of ultra-thin Nb films*
Nb superconducting films used in this study were deposited onto $SiO_2$/Si (100) or *R*-plane sapphire substrates by a home-made DC-magnetron sputtering system. The base pressure of the system was in the range of $10^{-9}$ mbar. Before film growth, substrates were baked inside the sputtering chamber at about 400 °C for 3 hours to degas. Nb with purity≥99.95% was used as the target, and Nb films were deposited on the substrates at room temperature. Since Nb is an active metal and its superconducting properties could be degraded by even very small amount of impurity gas such as $O_2$, a high purity Ar gas (better than 99.999%) was supplied through a customized built-in Non-Evaporable Getter (NEG) [23] purifier before flowing into the sputtering chamber. Sputtering parameters are listed in table 1. The deposition rate was deduced from film thickness and deposition time. The film thickness was determined by several methods, including AFM, SEM cross-sectional graph and X-ray reflectivity (XRR). The XRR results measured on ultrathin Nb films are shown in figure 1. Experimental data of all the specimens were fitted. Nb film thickness from 3.44 nm to 8.94 nm with about 0.5 nm surface roughness and 2-3 nm $Nb_2O_5$ over-layer were deduced from the fitting.

**Table 1.** Sputtering Parameters for Nb films.

| Conditions | Nb film |
|---|---|
| Base pressure | <$5\times10^{-9}$ mbar |
| Ar pressure | $4.9\times10^{-3}$ mbar |
| Power | 208 W |
| Deposition rate | 0.56 nm/sec |

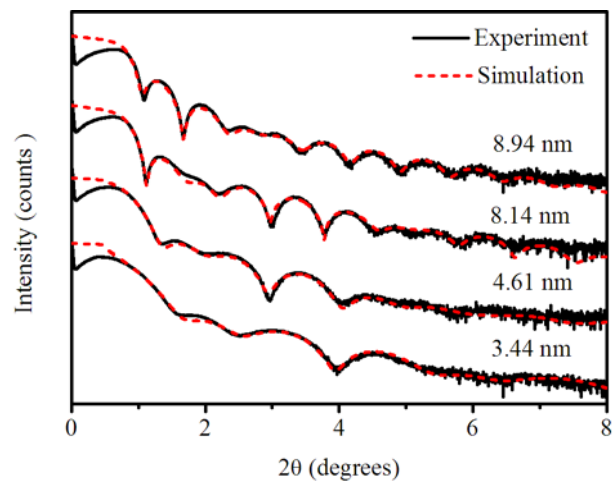

**Figure 1.** X-ray reflectivity (XRR) measurements performed on ultra-thin Nb films. Experimental results (black solid line) of all the specimens are in agreement with the results of the theoretical simulation (short dash), from where the thickness of Nb films from 3.44 nm to 8.94nm and 2-3 nm $Nb_2O_5$ over-layer on Nb are deduced.

*2.2. NIL mold fabrication*

*2.2.1. NIL hard mold.* The imprint hard mold based on Si wafer was fabricated by using a 2-step process, as illustrated in figure 2. First, alignment marks and meander lines with a width of about 85 nm covering a 10×10 μm² area were patterned at the central part of Si wafer by means of EBL with PMMA (positive tone) resist and RIE process based on a $CF_4/O_2$ gas mixture which selectively etched away all the unwanted Si material. The Si wafer was spin coated with 3% PMMA 495 electron beam resist. A sensible balance between achieving higher filling factor in EBL process and higher aspect radio in RIE, the resist thickness was optimized to be ~ 100 nm to fabricate meander patterns. The nanowires were then written at 20 kV by Raith 150 EBL system with an area dose of 260 μC/cm$^2$, and were transferred 55 nm down into Si by RIE. In the second step, a large square terrace pattern of S1813 (positive tone) photoresist (50×50 μm$^2$) was defined by UV-photolithography process, which was used to protect the meander area. RIE was then used to produce the surface relief structures on Si required for NIL. Figure 3 shows SEM images of the fabricated Si mold with meander-type lines.

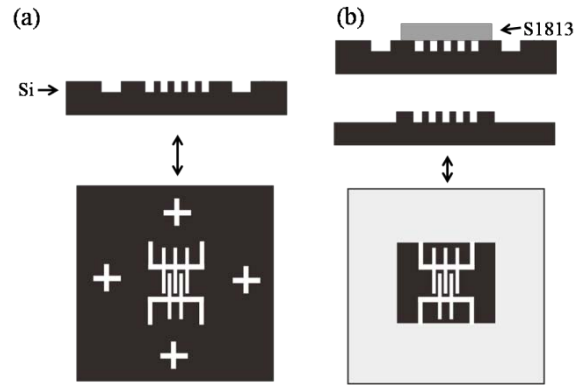

**Figure 2.** Schematic of the fabrication of NIL hard mold based on Si. (a) Meander lines with a width of about 85 nm covering a 10×10 μm$^2$ area were patterned at the central part of Si wafer by means of EBL with PMMA (positive tone) resist and RIE process based on a CF$_4$/O$_2$ gas mixture to transfer 55 nm down into Si. (b) A large square terrace pattern (50×50 μm$^2$) of S1813 (positive tone) photoresist was defined by UV-photolithography process, which was used to protect the meander area. RIE was then used to produce the surface relief surface relief structures on Si required for NIL

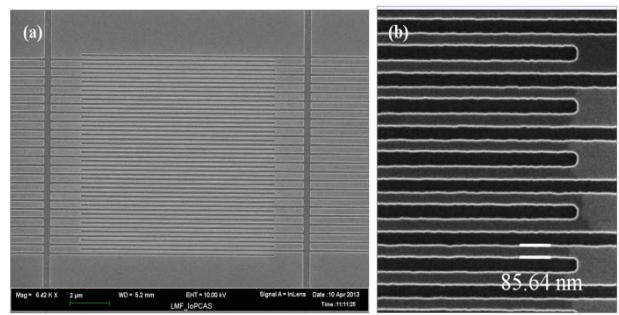

**Figure 3.** (a) The SEM image of the Si meander nanowire. (b) shows the SEM image of the enlarged Si meander of (a). The width of meander line is 85.64 nm and the filling factor of about 50%

*2.2.2. NIL soft mold.* Thermal-NIL was exploited to define the soft mold by Eitre® 3 Nano-Imprinter system from Obducat. Unlike in conventional thermal-NIL, where thermal imprint resist is used, here IPS acted as both the resist and the substrate. IPS is a transparent polymeric thermoplastic material with a thickness of about 200 μm, which has great strength and toughness at room temperature, whereas it becomes highly elastic above 120 $^o$C. It can be cut into different shapes and sizes according to the hard mold, and used in NIL process after lifting off its proof film. The hard mold was pressed with its patterned side to IPS at 160 $^{o}$C and 40 bar for 1 minute to transfer the patterns from the hard mold to the IPS soft mold. During the imprint process, IPS has no damage to the Si hard mold. Therefore by adopting a soft mold the lifetime of the NIL hard mold can be largely enhanced. Moreover, the transparency of the IPS soft mold is crucial for the following UV-NIL process, which is a chemical process thus can be carried out at lower temperature and pressure.

*2.3. SNSPD fabrication*

*2.3.1. Fabrication of meander lines.* UV-NIL and RIE were used to transfer the patterns from the IPS soft mold to Nb film, as illustrated in figure 4. The fabrication details are as follows. The Nb film of about 4 nm thick was spin coated with UV curable resist TU2-60 (the thickness of the resist is about 60 nm when the resist spinner rotates 3000 r/min) and was pre-baked at 95 $^{o}$C for 3 minutes to evaporate the organic solvent in the resist. The fabricated soft mold was then positioned above the Nb film, allowing the pattern side of the IPS and TU2 to confirm to each other. UV light was applied through the transparent IPS to cure the TU2 resist on the Nb films at 80 $^{o}$C and 30 bar for 5 minutes. The complete conditions of the SNSPD fabrication using NIL are listed in table 2. After lifting off the soft mold, TU2 with protrusion features was formed. IPS is a good non-adhesive material due to the fact that it contains fluorine, therefore it is prone to mold release. The depth of the curved resist relies on the depth of pattern on the mold, the thickness of the resist, and other imprint conditions (temperature, pressure and time of UV light exposure). After the imprint process, the patterned resist layer was used directly as an etching mask in the subsequent RIE step based on a $SF_6$/Ar gas mixture to remove the residue resist and unwanted Nb, to form meander-type nanowire. By UV-NIL, the time consumption of fabrication of a large detecting area is identical to that of a small area, which promises high-throughput patterning of nanostructures. The SEM images of a fabricated Nb meander line with the width and the filling factor of about 90 nm and 50% respectively are shown in figure 5. The width of line and the filling factor based on Nb generated are no noticeable differences from that of the imprint mold based on Si.

**Table 2.** Nanoimprint conditions.

|  | Temperature ($^{o}$C) | Pressure (Bar) | Time (s) | UV exposure |
|---|---|---|---|---|
| Thermal NIL | 160 | 40 | 60 | no |
|  | 100 | 40 | 20 | no |
| UV-NIL | 80 | 30 | 240 | no |
|  | 80 | 30 | 300 | yes |
|  | 80 | 30 | 120 | no |

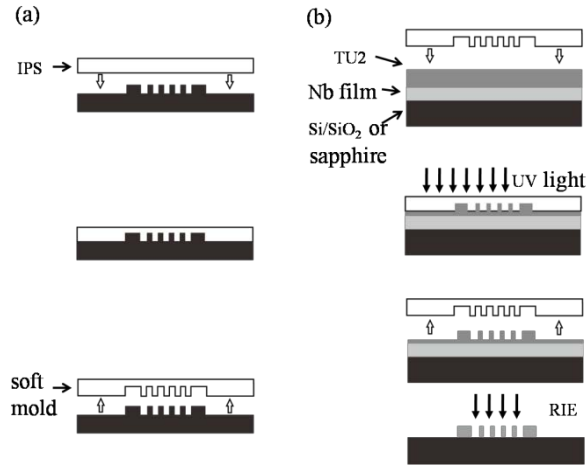

**Figure 4.** Schematic of the fabrication of meander lines based on ultra-thin Nb films. (a) The soft mold was defined by hard mold using thermal NIL. (b) By UV-NIL meander, patterns on TU2 risist were formed. RIE based on a $SF_6$/Ar gas mixture was then used to remove the residue resist and unwanted Nb, to form the meander-type nanostructure.

*2.3.2. Fabrication of coplanar waveguide.* In application of meander nanowires as SNSPDs, in order to allow a high bandwidth connection to room temperature readout electronics, the meander lines were coupled through a 50 Ω -Ti/Au- 5/100 nm coplanar waveguides which were fabricated by UV lithography, e-beam evaporation followed by the lift-off technique. In addition, Nb detectors are significantly more susceptible than NbN to thermal instability (latching) at high bias. Nevertheless, the detectors could be stabilized by reducing the input resistance of the readout circuit [7]. Taking this into consideration, an about 20 Ω -Ti/Au -5/100 nm meander-type resistance which is near and parallel to the superconducting nanowire was fabricated together with the coplanar waveguide, as also illustrated in figure 5.

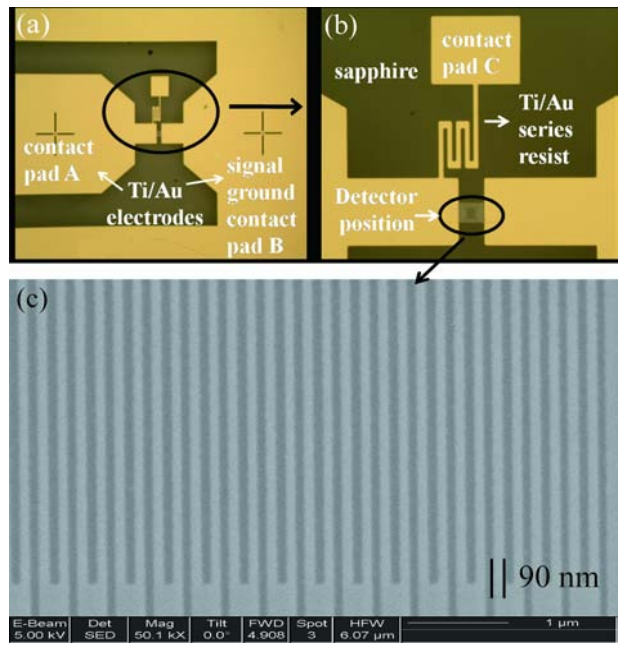

**Figure 5.** (a) Optical micrograph of an SNSPD detector chip. The central part outlined in (a), including the Ti/Au contact pads, a parallel resistor of about 20 Ω and the meander nanowire, is magnified in (b). The resistor bridges pad A and C, and the outer pad B is grounding electrode. The superconducting characteristics of nanowires were measured by bonding the electrode lead at pad A. When the meander nanowire is used as SNSPD, the Ti/Au resistor will be connected in parallel with with the detector by shorting pad B and C. (c) shows the SEM image of the enlarged Nb meander by NIL process. The mean width of superconducting stripes is about 90 nm. The etched portion is dark.

## 3. Superconducting properties

The superconducting critical temperatures ($T_c$) of Nb films with the thickness from 3 nm to 100 nm on *R*-sapphire substrates were studied in PPMS (physical property measurement system). Figure 6 shows $T_c$ and the width of transition ($\Delta T$) dependence of the thickness of Nb films (*t*). When $t > 10$ nm $T_c$ decreases and $\Delta T$ increases gradually with decreasing of *t*. However, a dramatic drop of $T_c$ and raced up of $\Delta T$ with reducing of *t* occur when $t < 10$ nm which meet the requirement of SNSPDs. Therefore, the choice of the Nb film thickness for fabricating meander nanowires need to balance the superconducting properties and the basic working mechanism of SNSPDs. Nb films of about 4 nm thick with $T_c \sim 4.65$ K were used to pattern meander-type structures in our experiments.

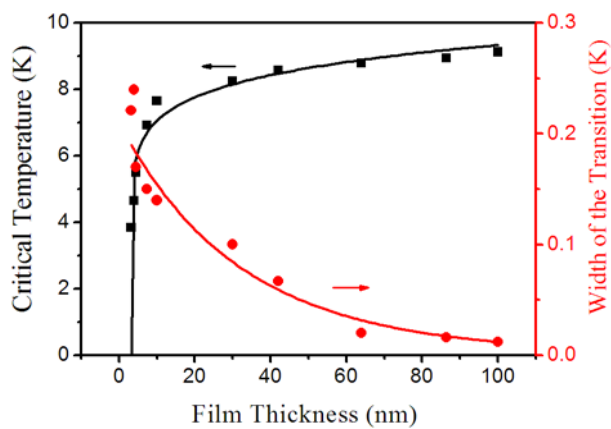

**Figure 6.** The superconducting critical temperature ($T_c$) and the width of transition ($\Delta T$) dependence of the Nb film thickness (*t*) on sapphire substrates. The lines are to guide the eye.

The temperature dependence of resistance was measured for the meander structures made on sapphire substrates by both EBL and NIL techniques. The width was about 90 nm and the filling factor was about 50%. The results are shown in figure 7. The resistive transition of the meander nanowire by means of NIL technique exhibits a $T_c$ around 4.3 K. As compared with the virgin film, the critical temperature of reactive-ion-etched nanowires (both by EBL and NIL) was reduced only slightly. However, $T_c$ values of the specimens patterned by NIL and EBL are almost the same, which demonstrates that NIL is a credible alternative to EBL in the fabrication of SNSPDs, especially for detectors with large areas.

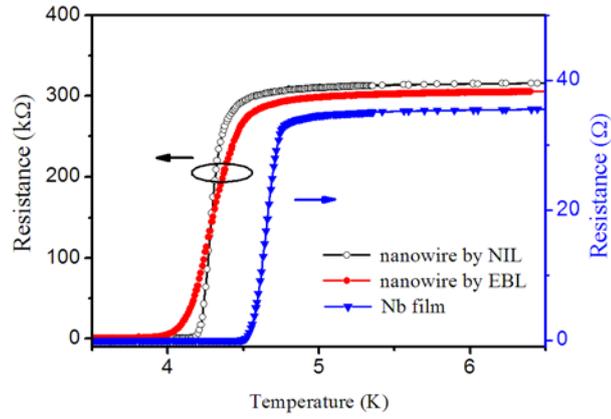

**Figure 7.** Resistance versus temperature (*R–T*) dependencies of the virgin Nb film (4nm thick) (blue triangles), meander lines by EBL (red circles), and ones by NIL (black open circles) based on sapphire. The specimen patterned by NIL exhibits a $T_c$ around 4.3 K.

The meander lines of about 90 nm wide covering a 10×10 μm² area with different filling factors (*f* = 50%, 45% and 33%) were also fabricated by NIL process using different Si hard mold. As shown in figure 8, the $T_c$ (4.5 K) of the sample with *f* = 33% is better than that of other samples. Because the defects could constrict the superconducting properties of the samples, we believe that the high Tc value observed in the *f* = 33% is probably due to the shorter total length of the meander, and therefore the reduced number of defects along the strip.

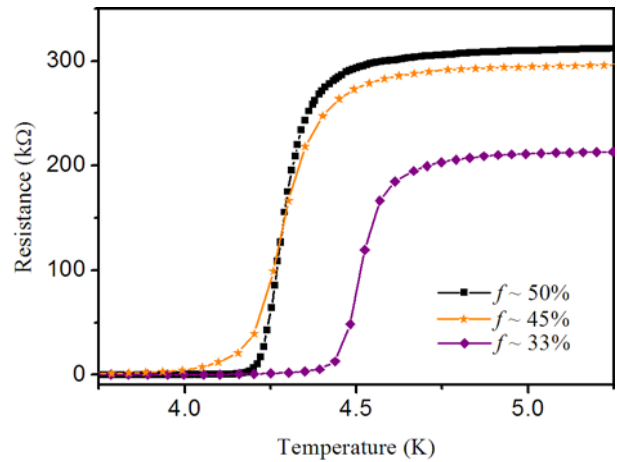

**Figure 8**. Resistance versus temperature (*R–T*) dependency of the samples with different filling factors (*f*). The nanowires of three samples are all about 90 nm.

The current-voltage (*I–V*) characteristic curve of the meander wire with 50% filling factor was measured in the current-bias mode at 2 K in the same experimental setup. The *I–V* curve shows a hysteresis behavior and a sharp jump from superconducting to normal state at $I_c$ of around 1.5 μA, as shown in Figure 9. The critical current density ($J_c$) evaluated from *I–V* curve is about $4.2 \times 10^5$ A/cm². Further optimization of the NIL processes is expected to yield meander nanowires with higher $J_c$.

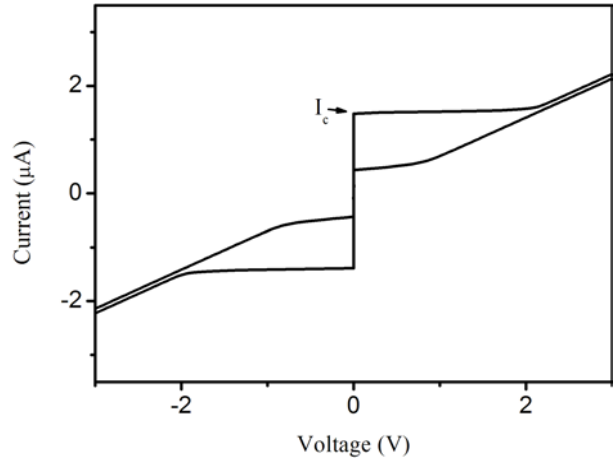

**Figure 9.** Current-voltage characteristic of the meander wire with 50% filling factor by NIL at 2 K. $I_c$ is about 1.5 μA, and $J_c$ evaluated from $I$–$V$ curve is about $4.2 \times 10^5$ A/cm$^2$.

## 4. Conclusion

In summary, we have demonstrated a parallel fabrication technique, a combination of thermal-NIL and UV-NIL, to fabricate meander-type devices based on ultra-thin Nb films. Meander lines of about 90 nm width with different filing factors covering effective detection areas of 10×10 μm$^2$ have been successfully fabricated. Patterned by the NIL technique, the meander lines are uniform and exhibit little degradation in superconducting properties. We believe that the devices fabricated by the NIL technique are suitable for single phonon detection.

Moreover, this technique can also be used for making meander-type SNSPDs on superconducting thin films of other materials, such as NbN, NbTiN and WSi. SNSPDs containing arrays of meander-type nanowires could provide even higher counting rate and photon-number resolution. For this kind of multi-element devices, the fabrication process using NIL is a more desirable because of the high-throughput and low cost.

**Acknowledgements**

This work is supported by the National Basic Research Program of China (973 Program, No. 2011CBA00106), as well as by National Natural Science Foundation of China (No. 11104333 and No. 10974243). The authors thank Wei Peng for technical assistance with X-ray reflectivity, Aizi Jin for helpful discussion of NIL technique.

**References**

[1] Hadfield R H 2009 Single-photon detectors for optical quantum information applications *Nat. Photonics* **3** 696-705
[2] Gol'tsman G N, Okunev O, Chulkova G, Lipatov A, Semenov A, Smirnov K, Voronov B, Dzardanov A, Williams C, and Sobolewski R, 2001 Picosecond superconducting single-photon optical detector *Appl. Phys. Lett.* **79** 705–7
[3] Semenov A D, Gol'tsman G N, and Korneev A A 2001 Quantum detection by current carrying superconducting film *Physica* C **351** 349–56

[4] Natarajan C M, Tanner M G, and Hadfield R H 2012 Superconducting nanowire single-photon detectors: physics and applications *Supercond. Sci. Technol.* **25** 063001

[5] Annunziata A J, Frydman A, Reese M O, Frunzio L, Rooks M, and Prober D E 2006 Superconducting Nb nanowire single photon detectors *Advanced Photon Counting Techniques* **6372** 63720V

[6] Fujii G, Fukuda D, Numata T, Yoshizawa A, Tsuchida H, Inoue S, and Zama T 2009 Fiber Coupled Single Photon Detector with Niobium Superconducting Nanowire *QuantumCom* LNICST **36** 220–4

[7] Annunziata A J, Santavicca D F, Chudow J D, Frunzio L, Rooks M J, Frydman A, and Prober D E 2009 Niobium Superconducting Nanowire Single-Photon Detectors *IEEE T APPL SUPERCON* **19** 327-31

[8] Gol'tsman G N *et al* 2001 Ultrafast superconducting single-photon detector *Journal of Modern Optics* **56** 1970-80

[9] Miki S, Yamashita T, Terai H and Wang Z 2013 High performance fiber-coupled NbTiN superconducting nanowire single photon detectors with Gifford-McMahon cryocooler *Opt. Express* **21** 10208-14

[10] Marsili F *et al* 2013 Detecting single infrared photons with 93% system efficiency *Nat. Photonics* **7** 210-4

[11] Hu X L, Zhong T, White J E, Dauler E A, Najafi F, Herder C H, Wong F N C, and Berggren K K 2009 Fiber-coupled nanowire photon counter at 1550 nm with 24% system detection efficiency *Opt. Lett.* **34** 3607-9

[12] Dauler E A, Robinson B S, Kerman A J, Yang J K W, Rosfjord K M, Anant V, Voronov B, Gol'tsman G N, and Berggren K K 2007 Multi-Element Superconducting Nanowire Single-Photon Detector *IEEE T APPL SUPERCON* **17** 279-84

[13] Bachar G, Baskin I, Shtempluck O, and Buks E 2012 Superconducting nanowire single photon detectors on-fiber *Appl. Phys. Lett.* **101** 262601

[14] Delacour C 2007 Superconducting single photon detectors made by local oxidation with an atomic force microscope *Appl. Phys. Lett.* **90** 191116

[15] Yang X Y, You L X, Wang X, Zhang L B, Kang L and Wu P H 2009 Local anodic oxidation of superconducting NbN thin films by an atomic force microscope *Supercond. Sci. Technol.* **22** 125027

[16] Troeman A G P, Derking H, Borger B, Pleikies J, Veldhuis D, and Hilgenkamp H 2007 NanoSQUIDs based on Niobium constrictions *Nano lett.* **7** 2152-6

[17] Chou S Y, Krauss P R, and Renstrom P J 1996 Nanoimprint lithography *J. Vac. Sci. Technol.* B **14** 4129-33

[18] L. Jay Guo 2007 Nanoimprint lithography: methods and material requirements *Adv. Mater.* **19** 495–513

[19] Bergmair I *et al* 2012 Nano- and microstructuring of graphene using UV-NIL *Nanotechnology* **23** 335301

[20]　Cheng X, Li D W and Guo L J 2006 A hybridmask–mould lithography scheme and its application in nanoscale organic thin film transistors *Nanotechnology* **17** 927–32

[21]　Austin M D, Ge H X, Wu W, Li M T, Yu Z N, Wasserman D, Lyon S A, and Chou S Y 2004 Fabrication of 5 nm linewidth and 14 nm pitch features by nanoimprint lithography *Appl. Phys. Lett.* **84** 5299-301

[22]　Xia Y N and Whitesides G M 1998 Soft lithography *Rev. Mater. Sci.* **28** 153-84

[23]　Jin Y R, Song X H and Zhang D L 2009 Grain-size dependence of superconductivity in dc sputtered Nb films *Sci. China Ser* G **52** 1289-92